# Role of Se vacancies on Shubnikov de Haas oscillations in $Bi_2Se_3$: a combined magneto-resistance and positron annihilation study

T. R. Devidas, E. P. Amaladass, Shilpam Sharma, R. Rajaraman, D. Sornadurai, N. Subramanian, Awadhesh Mani, C. S. Sundar and A. Bharathi[a]

*Materials Science Group, Indira Gandhi Centre for Atomic Research, Kalpakkam - 603102, Tamil Nadu, India*

**Abstract** – Magneto resistance measurements coupled with positron lifetime measurements, to characterize the vacancy type defects, have been carried out on the topological insulator (TI) system $Bi_2Se_3$, of varying Se/Bi ratio. Pronounced Shubnikov de Haas (SdH) oscillations are seen in nominal $Bi_2Se_{3.1}$ crystals for measurements performed in magnetic fields up to 15 T in the 4 K to 10 K temperature range, with field applied perpendicular to the (001) plane of the crystal. The quantum oscillations, characteristic of 2D electronic structure, are seen only in the crystals that have a lower concentration of Se vacancies, as inferred from positron annihilation spectroscopy.

**Introduction.** – Topological insulators (TI), having an insulating bulk state and a spin polarized metallic surface state with a unique Dirac cone dispersion, have been extensively investigated in the last few years both theoretically and experimentally [1-5]. The interesting dissipation-less transport behavior of TI, stems from the fact that the conducting states have their spin locked to the momentum, and are as a consequence protected from backscattering. Although the characterization of the electronic structure of topological insulators has been accomplished by Angle Resolved Photoemission Spectroscopy (ARPES) and Scanning Tunneling Microscopy (STM), their signature in transport behavior is of interest and would be crucial for applications [4, 5]. An important theme in the research of TI materials [2, 3] has been to investigate methods to reduce unintentionally-doped bulk carriers that hinder observations of interesting 2D transport properties. The presence of vacancy type defects and anti-site defects influence significantly the electronic properties of the Topological insulators.

$Bi_2Se_3$, an extensively investigated TI system, has a tetradymite structure [Space group $R\bar{3}m$], consisting of quintuple layers [Se(1)-Bi-Se(2)-Bi-Se(1)] separated by Van der Waals gap (see Fig. 1(a)). This structure allows for three types of vacancies, viz., the Se(1) vacancy in the Selenium layer closest to the Van der Waals gap, the Se(2) vacancy in the Selenium layer inside the quintuple layer and one at the Bismuth atom of the structure. In addition, there are anti-site defects. First principles electronic structure calculations have been extensively carried out to investigate the energetics of the formation of various point defects under various growth conditions [6-10], and their influence on bulk conductivity. The vacancy at the Se site, $V_{Se}$ and the anti site defects viz., $Se_{Bi}$ and $Bi_{Se}$ are the intrinsic defects with low formation energies that account for the observed n- doped behavior of $Bi_2Se_3$. On the surface, Se vacancies are usually seen as clear triangular depressions in STM experiments [9, 11], and High Resolution Transmission Electron Microscopy studies have been carried out [12] to provide evidence for the anti-site defects in $Bi_2Se_3$.

In the present investigation, we have used positron annihilation spectroscopy [13, 14], coupled with theoretical calculations of annihilation characteristics to investigate vacancy type defects in $Bi_2Se_3$, and correlate with the magneto-resistance behaviour in three crystals with varying Se/Bi ratio. It is shown that positrons are sensitive to the presence of Se(1) vacancies, whose concentration has been estimated. Magneto-resistance measurements indicating quantum oscillations characteristic of 2D electronic structure [2], are seen only in the crystals that have a lower concentration of Se(1) vacancies.

**Experimental.** – Single crystals of $Bi_2Se_3$ with nominal compositions $Bi_2Se_3$, $Bi_2Se_{3.1}$, $Bi_{2.1}Se_3$, that correspond to stoichiometric, Se-rich and Bi-rich samples, were synthesized by melting the required quantities of high purity elements in evacuated quartz tubes at 850ºC (for 24 hours) followed by slow cooling at the rate of 1ºC/hour until 550ºC, keeping at this temperature for 24 hours followed by rapid cooling. The obtained crystals could be easily cleaved along the basal plane leaving a silvery shiny mirror like surface. Laue diffraction measurements were carried out on the cleaved crystals using a Molybdenum x-ray source, and the patterns were recorded in transmission mode using a HD-CR-35 NDT image plate system. Powder x-ray diffraction (XRD) was carried out on powdered single crystals in an APD 2000 PRO (GNR Analytical Instruments Group, Italy.) diffractometer in the Bragg-Brentano geometry.

[a]Author to whom any correspondence should be addressed
[a]E -mail: `bharathi@igcar.gov.in`

Temperature dependent resistivity measurements, were carried out in a dipper cryostat in the Van der Pauw geometry, with contacts made on the (001) face of the crystals. Magneto-transport measurements were carried out in both, the Van der Pauw geometry and linear geometry in a commercial, 15 T Cryogen-free system from Cryogenic Ltd., UK. The Hall measurements were obtained in the Hall bar geometry in the same setup. Both the resistivity and Hall Effect measurements were carried out simultaneously on each of the single crystals. The contacts to the samples, placed in a puck holder, were done with 25 micron Au wires, with Ag paste, that cures at room temperature.

The investigation of vacancy defects in samples have been carried out using positron lifetime measurements. For positron lifetime measurements, the $Na^{22}$ positron source evaporated on 1.25 micron Ni foil was sandwiched between two crystal surfaces. The measurements were carried out using a fast-fast coincidence spectrometer, having a time resolution of ~260 ps [15]. The positron data were analyzed using LT program [16]. The results of positron lifetime measurements have been analysed in conjunction with the theoretical calculations of positron density distribution and lifetimes in perfect crystal and those containing vacancy defects.

**Results and Discussions.** – The powder XRD patterns for $Bi_2Se_3$, $Bi_2Se_{3.1}$ and $Bi_{2.1}Se_3$ single crystals are shown in Fig. 1(b). All the major peaks could be indexed to those of the Rhombohedral $Bi_2Se_3$ structure and no Se or Bi impurity peaks are observed in the patterns. The c-axis lattice parameters obtained from the three samples viz., $Bi_{2.1}Se_3$, $Bi_2Se_3$ and $Bi_2Se_{3.1}$ are respectively 28.62 $\pm$ 0.003 Å, 28.61 $\pm$ 0.005 Å and 28.65 $\pm$ 0.001 Å (see Table 1). These results are consistent with the results of Huang et. al [12]. The Laue diffraction patterns measured in the transmission geometry on (001) cleaved crystals (confirmed by back-reflection Laue) are shown in Fig. 1(c). It is noted that while the spots are sharp in the $Bi_2Se_3$ sample, predominant radial streaking is seen in the non-stoichiometric sample $Bi_2Se_{3.1}$, pointing to the presence of lattice deformations. These could manifest due to stacking faults along the *c*-axis and strain inhomogeneities [17, 18, 19]. In addition some spots show streaking in directions along the Debye arc, indicative of the presence of small angle tilt grain boundaries [18, 19]. The Laue pattern of the Bi excess, $Bi_{2.1}Se_3$ crystal has larger spots, probably due to uniform strain due to excess Bi. The resistivity versus temperature measured in the Van der Pauw geometry for all the three samples is shown in Fig. 2. The positive temperature co-efficient of resistivity indicates metallic behaviour in all samples, with a tendency to saturate below 30 K. In the case of Se excess sample, $Bi_2Se_{3.1}$, there is a marginal increase in resistivity below ~30 K, similar to earlier results[20]. Such an up-turn, points to the significant contribution to the conductivity from the thermally activated carriers from impurity bands vis-à-vis the band conductivity, and it is in this system with low carrier density clear magneto-resistance oscillations are seen, as discussed below (cf. Figs. 3 and 4).

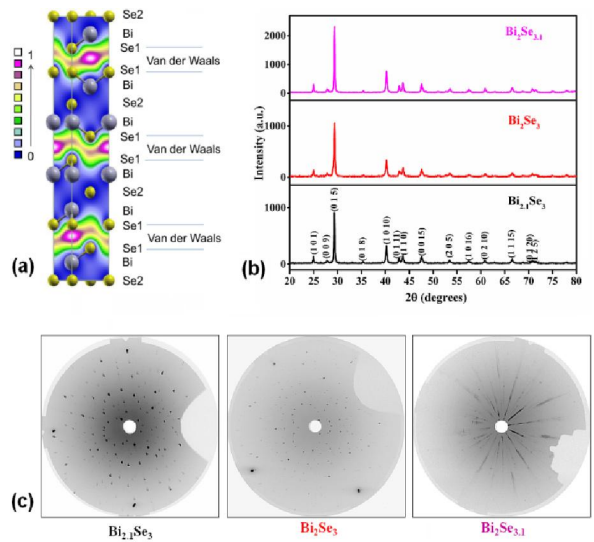

Fig. 1: (Colour on-line) (a) The (110) plane of the $Bi_2Se_3$ structure, showing the important structural features. The positron density distribution in the unit cell shows that the positron samples the Van der Waal gap between the quintuplet layers. (b) Powder X-ray diffraction pattern of the three samples of varying Se/Bi ratio, used in the present study (c) Forward scattering Laue patterns of the corresponding crystals.

To get an idea of the nature of charge carriers, Hall measurements were carried out at 4.2 K, which are shown in the insets of Fig. 2. It is inferred that the carriers are electrons and the carrier density for the three samples, estimated from the slope of the Hall data at low fields, are indicated in Table 1. The corresponding carrier mobility ($\mu$) values at zero field obtained from the measured conductivity $\sigma$ and carrier density (n), using the formula $\mu = \sigma/ne$, is also shown in Table 1.

The results of magneto-resistance measurements, carried out in the magnetic field range of -15 T to 15 T at 4.2 K in the Van der Pauw geometry are shown in Fig. 3. It is seen from the figure that clear quantum oscillations are observed in the Se excess $Bi_2Se_{3.1}$ sample, but not so clearly in the other compositions. The Fourier transform of the ρ versus 1/B (see insets) indicates that only in the sample with Se excess, the Shubnikov de Haas (SdH) oscillations is dominated by a single frequency of 46.90 T.

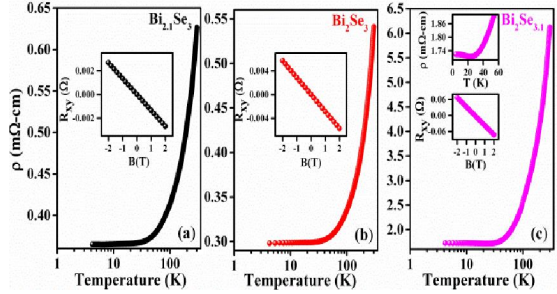

Fig. 2: (Colour on-line) Resistivity measurement on (a) $Bi_{2.1}Se_3$, (b) $Bi_2Se_3$ and (c) $Bi_2Se_{3.1}$ single crystals. The inset shows the Hall resistance versus magnetic field at 4.2 K. In the case of $Bi_2Se_{3.1}$ sample, there is a small upturn at low temperature as shown in the inset.

To analyse the magneto-resistance behavior in $Bi_2Se_{3.1}$ in detail, the magneto-transport measurements on the (001) oriented single crystals were carried out, in the linear geometry at three different temperatures of 4.2 K, 7 K, 10 K in the -15 T to +15 T magnetic field range and these results are shown in Fig. 4(a).

The oscillatory component of resistivity $\Delta\rho_{xx}$, obtained from measured longitudinal resistance, has been analysed in terms of the Lifshitz-Kosevich (L-K) equation [21, 22]:

$$\Delta\rho_{xx} = A_0 R_T R_D R_S \cos\left[2\pi\left(\frac{F}{B} + \frac{1}{2} + \beta\right)\right] \quad (1)$$

Where

$$A_0 = \left(\frac{\hbar\omega_c}{2\varepsilon_F}\right)^{\frac{1}{2}}; R_T = \frac{\left(2\pi^2 k_B T/\hbar\omega_c\right)}{\sinh\left[2\pi^2 k_B T/\hbar\omega_c\right]}; R_D = e^{-\left[\frac{2\pi^2 k_B T_D}{\hbar\omega_c}\right]}$$

$$R_S = \cos\left(\frac{1}{2}\pi g \frac{m_0}{m^*}\right)$$

Table 1: Lattice parameters from XRD, carrier concentration n and mobility µ, from low field Hall data and positron lifetimes in $Bi_{2.1}Se_3$, $Bi_2Se_3$ and $Bi_2Se_{3.1}$ single crystals. The last column indicates the Se(1) vacancy concentration $V_{Se}$, as obtained from positron experiments.

| System | a (Å) | c (Å) | n (cm$^{-3}$) | µ (cm$^2$/V-s) | Positron Lifetime τ (ps) | $V_{Se}$ (cm$^{-3}$) |
|---|---|---|---|---|---|---|
| $Bi_{2.1}Se_3$ | 4.1420±0.006 | 28.62±0.003 | 2.83*10$^{19}$ | 604 | 229±2 | 5.41*10$^{17}$ |
| $Bi_2Se_3$ | 4.1405±0.0008 | 28.61±0.005 | 4*10$^{19}$ | 260 | 224±2 | 2.86*10$^{17}$ |
| $Bi_2Se_{3.1}$ | 4.1419±0.0002 | 28.65±0.001 | 6.14*10$^{18}$ | 3635 | 216±2 | 1.1*10$^{17}$ |

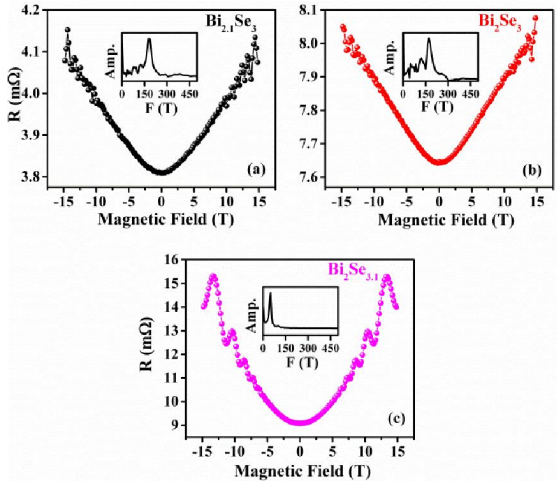

Fig. 3: (Colour on-line) Magneto-resistance at 4.2 K in (a) $Bi_{2.1}Se_3$, (b) $Bi_2Se_3$ and (c) $Bi_2Se_{3.1}$. Quantum oscillations in the magneto-resistance with a single frequency are clearly seen only for the Se rich $Bi_2Se_{3.1}$ sample; insets - the Fourier transform of the SdH oscillations

The frequency of SdH oscillations for 2D systems is given by $F = (4\pi^2 \hbar n_s)/e$; where $n_s = (k_F^2)/4\pi$ (per spin) is the 2D carrier density [23]. Excellent fits to the above eq. were obtained for the $\Delta\rho_{xx}$ versus 1/B data, at all temperatures. A representative fit of the resistivity data obtained at 4.2 K, in the $Bi_2Se_{3.1}$ sample, to L-K eq. is shown in Fig. 4(b), the best fit is obtained for a single frequency F = 46.90 T (the same frequency as also obtained from the Fourier transform of the oscillations) and the phase factor β = 0.62. The values of the fitted parameters are; Fermi wave-vector $k_F$ = 3.749*10$^6$ cm$^{-1}$; Dingle Temperature $T_D$ = 7.32 K, Cyclotron mass $m^*$ = 0.181$m_o$ and g = 2. From the above fitted parameters, the following physical quantities have been estimated, viz., the quantum scattering time $\tau_q = \hbar/(2\pi k_B T_D)$ = 0.166 ps, Fermi energy $\varepsilon_F = (\hbar^2 k_F^2)/(2m^*)$ = 0.029 eV, Fermi velocity $v_F = (\hbar k_F)/m^*$ = 2.42*10$^7$ cms$^{-1}$, 2D carrier density $n_s$ = 1.12*10$^{12}$ cm$^{-2}$, Surface mean free path $l_s^{SdH} = v_F \tau_q$ = 40±2 nm and surface mobility $\mu_s^{SdH} = (e l_s^{SdH})/(\hbar k_F)$ = 1619 cm$^2$/V-s. The parameters obtained from the above analysis are in agreement with earlier results in the $Bi_2Se_3$ system [20], and it is also noted that the surface carrier density $n_s$ is much smaller than the bulk carrier concentration n (see Table 1).

It is well known that in the case of the 3D topological insulators, the transport has contributions both from the 2D and 3D electronic states, and as indicated by Ando [2], the analysis of the phase factor (Berry's phase) of the SdH oscillations, helps provide unambiguous evidence if the transport arises from Dirac dispersion of electrons.

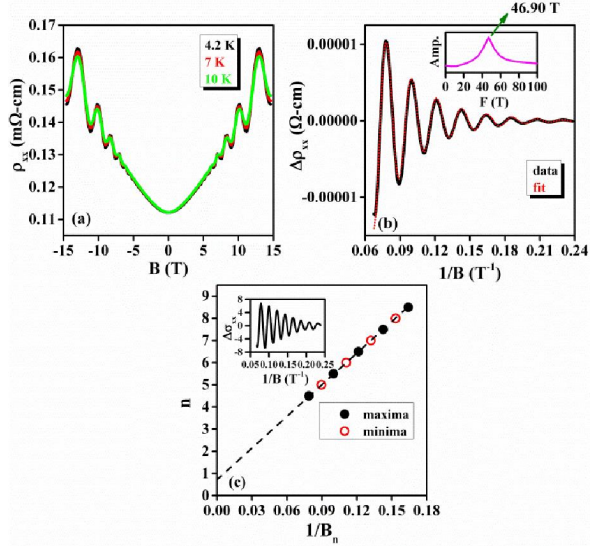

Fig. 4: (Colour on-line) (a) Magneto-resistivity of $Bi_2Se_{3.1}$ single crystal at various temperatures, (b) Representative fit of the Lifshitz-Kosevich equation to the SdH oscillations, obtained from $\Delta\rho_{xx}$ data at 4.2 K; inset shows the frequency of oscillations obtained from the Fourier transform of the oscillations, (c) The Landau level fan diagram - $1/B_n$ vs. n plot, where $B_n$ is the field at which maxima and minima are observed. Indices n and n+1/2 are assigned to minima and maxima in conductivity (inset) respectively. Straight line fit with a slope value fixed as the SdH oscillation frequency intercepts the index axis at 0.71

The data is converted in terms of conductivity using the relation $\sigma_{xx} = \rho_{xx}/(\rho_{xx}^2 + \rho_{xy}^2)$ to plot the Landau Level (LL) Fan diagram, wherein the minima and maxima in $\Delta\sigma_{xx}$ are assigned integers n and n+1/2 respectively and plotted with respect to $1/B_n$ as shown in Fig. 4(c). A linear extrapolation with the slope fixed by the oscillation frequency, resulted in the n-axis being intercepted at n = 0.71, as compared to the value of 0.62 obtained from the L-K equation fit, shown in Fig. 4(b). These values being close to 0.5, suggests that the SdH oscillations arise from Dirac electrons.

It is of interest to know as to why the SdH oscillations arising from a single frequency are seen only in the Se excess sample (cf. Fig. 3). In particular it is important to understand if this is in any way linked to Se vacancies that introduce charge carriers in the system, leading to shift of the Fermi level away from the Dirac point, and it is with this motivation, positron lifetime measurements have been carried out. Figure 5 shows the results of positron lifetime measurements carried out for the three compositions on which the magneto-resistance measurements have been carried out, viz., $Bi_{2.1}Se_3$, $Bi_2Se_3$ and $Bi_2Se_{3.1}$. The positron lifetime is observed to decrease systematically with the increase in Se/Bi ratio (see Table 2).

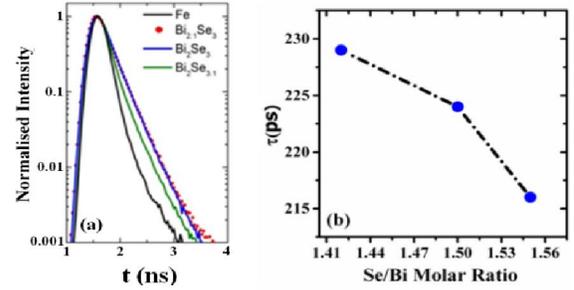

Fig. 5: (Colour on-line) (a) Raw positron lifetime data on $Bi_2Se_3$ single crystals as compared to that in Iron (Fe) (b) Positron lifetime value vs. Se/Bi molar ratio

Table 2: The calculated lifetimes in the perfect crystal of $Bi_2Se_3$ and at various vacancy defects. The positron binding energies at various vacancy defects is also shown. Se(1) vacancy in the Se layer closest to the Van der Waals gap, the Se(2) vacancy in the Se layer inside the quintuple layer.

| Species | $\tau$ (ps) | Binding energy |
|---|---|---|
| $Bi_2Se_3$(bulk) | 201 ps | 0.0 eV |
| Se(1) vacancy | 238 ps | 0.281 eV |
| Se(2) vacancy | 202 ps | 0.014 eV |
| Bi vacancy | 224 ps | 0.213 eV |

To understand the experimental positron data, theoretical estimates of positron lifetimes have been carried out from a calculation of the overlap of the positron density distribution with the electron density within the unit cell. These calculations have been carried out for both the perfect crystal, and crystal containing various vacancy defects, viz., vacancy at the Se(1), Se(2) and the Bi site. The positron density distribution has been calculated by solving the Schrödinger eq. for the positron, using the Doppler code [16], wherein the positron potential comprises of a repulsive Coulomb part from the ion core and an attractive part due to the superposed electronic density from the atoms. Fig. 6 shows the plot of the positron density distribution in crystalline $Bi_2Se_3$ and in the presence of a vacancy at the Bi, Se(1) and Se(2) sites. It is seen that in the perfect crystalline solid the positron density is confined to the Van der Waals gap in the $Bi_2Se_3$ structure, which are the open regions within the crystal. In the presence of Bi and Se(1) vacancy, it gets localized at the vacancy site. For the Se(2) vacancy within the quintuplet layer, the positron density is diffuse in that it samples both the vacancy and the Van der Waal gap. The above positron density distribution in the presence of vacancies has been calculated using a super-cell of 441 unit cells. Using the above positron density distribution, the annihilation rate is calculated from the overlap with the

T R Devidas *et al.*

electron density. The positron total annihilation rate ($\lambda_{tot}$) is given by contributions from the annihilation from core and valence electrons as [13], viz., $\lambda_{tot}=\lambda_{core}+\lambda_{val}$, where, $\lambda_{core}$ arises from the annihilations from the core electron density and $\lambda_{val}$ is the annihilation rate from the valence electron density. For these calculations, the electronic configuration of Bi is taken to be ([Xe] $4f^{14}$ $5d^{10}$) $6s^2$ $6p^3$ and Se to be ([Ar] $3d^{10}$ $4s^2$) $4p^4$, with the orbitals in brackets taken as core electrons and those outside as valence electrons. The above partitioning into valence and core electrons seems reasonable, as recent band structure calculations [10] suggest that the electronic structure of $Bi_2Se_3$ is well represented as $pp\sigma$ bonding of the 6p Bi and 4p Se orbitals. The annihilation rate was calculated taking into account the enhancement of valence electron density around the positron [24]. Using the above methodology, the positron lifetime in defect free $Bi_2Se_3$ has been calculated to be 201 ps. In the case of positron localization and annihilation at Se(1), Se(2) and Bi vacancies, the lifetimes are calculated to be 238 ps, 202 ps and 224 ps respectively (see Table 2). It is seen that the positron lifetime at Se(2) vacancy, within the quintuplet layer is not very different from that in the bulk. This is also reflected in the fact that the positron density distribution is not very localized for the Se(2) vacancy (cf. Fig. 6), and the binding energy is small as indicated in Table 2. However, the other Se vacancy, viz., the Se(1) vacancy near the Van der Waal gap is a deep trap for positrons with a lifetime substantially higher than that of the bulk. Experimentally, the positron lifetimes for the three compositions, viz., $Bi_{2.1}Se_3$, $Bi_2Se_3$ and $Bi_2Se_{3.1}$ have been measured to be 229 ps, 224 ps and 216 ps, respectively (cf. Table 1 and Fig. 5(b)). First it is noted that these are larger than the calculated bulk lifetime of 201 ps, clearly pointing to the localization and annihilation from vacancy type defects. To identify the specific vacancy defect, we first note that detailed first principles calculations [6], taking into account the spin orbit interactions, indicate that the formation energy of vacancy at the Bi site is significantly larger (by a factor of 3) than that at Se site, for both Bi rich and Se rich conditions, and it is the $V_{Se}$ that determines the n-doping level in $Bi_2Se_3$. Hence, we discount the vacancies at the Bi site as a possible reason for the observed increase in lifetime, compared to bulk. Amongst the two possible vacancy type defects at the Se site, we have already noted that the vacancy at Se(2) is a weak trap with a lifetime (202 ps), not different from the bulk value. Hence we infer that the positron lifetime in the $Bi_2Se_3$ samples arises due to positron trapping and annihilation at the Se(1) vacancy. Given this, from the measured value of lifetimes on samples of varying compositions (cf. Table 1), the concentration of Se(1) can be estimated, within the framework of the two state trapping model [13, 14]. The measured mean lifetime $\tau_m$ is given by $\tau_m = \tau_b ((1+K\tau_v))/((1+K\tau_b))$ where $\tau_b$ is the bulk lifetime (201 ps), $\tau_v$ is the lifetime at the Se(1) vacancy (238 ps) and K is the trapping rate. The trapping rate is defined as $K = \mu_v C_v$, where $\mu_v$ is the specific trapping rate at a vacancy and $C_v$ is the vacancy concentration. For the specific rate at the Se vacancy in $Bi_2Se_3$, we use the result [25] for trapping rate at Se vacancy in $CuInSe_2$, viz., $\mu_v = 1 \times 10^{15}$ $s^{-1}$. The resultant concentration of vacancies in the three samples is indicated in Table 1. It is seen that the concentration of vacancies (~ $10^{17}$), as obtained from positron experiments, is in accord with the theoretically calculated [6] vacancy concentration for the growth of $Bi_2Se_3$ under Se rich conditions. We further note (see Table 1) that the carrier density, as obtained from Hall measurements, is an order of magnitude larger than the concentration of $V_{Se}$. To understand this we take re-course to recent studies on thin films of $Bi_2Se_3$, which demonstrate with the help of calculations and Scanning Tunneling Microscopy data that the Dirac point, shifts, due to the introduction of low angle grain boundaries [19]. The presence of stacking faults and low angle grain boundaries introduced in the Se excess sample inferred from Fig. 1(c) can shift the Dirac point to lower energy as shown in schematic diagrams of the electronic structure, of stoichiometric and Se excess samples as in Fig. 7. Bulk Fermi surface is pinned to the bottom of the conduction band (CB)[26]. Band bending results due to equalization of surface (red line) and bulk (blue lines) Femi levels [26]. In the stoichiometric $Bi_2Se_3$, there is downward band bending, since the $E_F$ of surface state is higher than that of the bulk, whereas in the Se excess sample there is upward band bending, leading to a charge transfer from the bulk states to surface states resulting in the increased mobility seen in the Se excess sample (cf. Fig.7). To summarize, band bending [26] and shifting of Dirac point [19], lead to changes in electronic structure, that can account for the observed larger carrier density, not possible from the rigid band approach, using the observed Se vacancy concentration. Thus, the combined magneto-transport, positron measurements and Laue scattering studies as carried out in the present work demonstrates that defect characterization would play an important role in tuning the single crystal synthesis parameters, in order to control the disorder in the samples, which could be crucial in harnessing the interesting transport behaviour of the surface states in topological insulators.

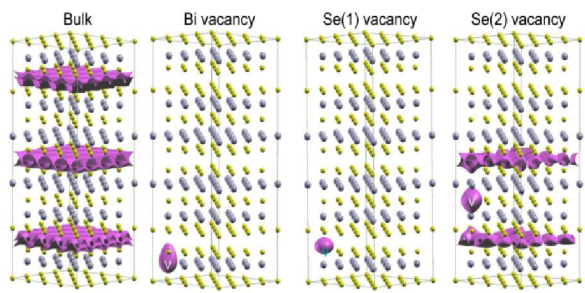

Fig. 6: (Colour on-line) Positron density distribution in crystalline $Bi_2Se_3$ and in the presence of vacancy at Bi, Se(1) and Se(2) sites. The positron density distribution for vacancies has been carried out for 441 super-cell, with vacancies in each layer. In defect free $Bi_2Se_3$, the positron density is confined to the Van der Waals gap, and in the presence of Bi and Se(1) vacancy, it gets localized at the vacancy site. For the Se(2) vacancy within the quintuplet layer, the positron samples both the vacancy and the Van der Waal gap. The overlap of this positron density distribution with the atomic superposition of electron density is used to calculate the corresponding positron lifetimes.

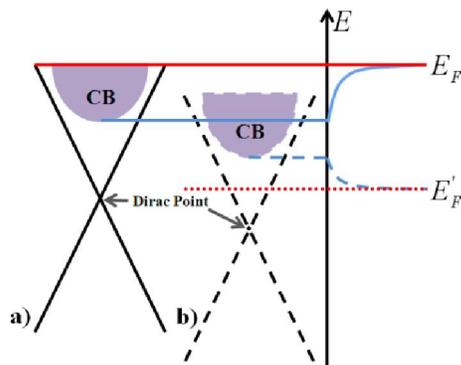

Fig. 7: (Colour on-line) (a) Downward band bending in $Bi_2Se_3$ (red lines denote the $E_F$ of the surface state, blue denote the Bulk $E_F$) [26] (b) in the Se excess sample (shown with broken line), Dirac point shifts due to strain leading to the upward band bending.

∗∗∗

**Acknowledgements.** – One of the authors, T. R. Devidas gratefully acknowledges the Department of Atomic Energy for providing the Senior Research Fellowship and Dr. V. Sridharan for the powder x-ray diffraction data on the various samples. The authors gratefully acknowledge UGC-DAE-CSR node at Kalpakkam for providing access to the 15 T cryogen free magneto-resistance facility. Prof. A. Thamizhavel (TIFR, Mumbai) is thanked for the Laue diffraction patterns on all the samples.